# New Approach to Policy Effectiveness for Covid-19 and Factors Influence Policy Effectiveness

Yile He[1, *]

[1]University of California, Davis, 95616, Davis, The United States
[*]Corresponding author. Email: abhe@ucdavis.edu

**ABSTRACT**
This study compared the effectiveness of COVID-19 control policies, including wearing masks, and the vaccine rates through proportional infection rate in 28 states of the United States using the eSIR model. The effective rate of policies was measured by the difference between the predicted daily infection proportion rate using the data before the policy and the actual daily infection proportion rate. The study suggests that both mask and vaccine policy had a significant impact on mitigating the pandemic. We further explored how different social factors influenced the effectiveness of a specific policy through the linear regression model. Out of 9 factors, the population density, number of hospital beds per 1000 people, and percent of the population over 65 are the most substantial factors on mask policy effectiveness, while public health funding per person, percent of immigration have the most significant influence on vaccine policy effectiveness. This study summarized the effectiveness of different policies and factors they associated with. It can be served as a reference for future covid-19 related policy.

***Keywords:*** *Covid-19, eSIR, Linear Regression.*

## 1. INTRODUCTION

COVID-19 is an infectious disease caused by SARS-CoV-2, which has been declared a global public health emergency. As of 28 October 2021, it has led to more than 45.7 million confirmed cases and resulted in more than 742 thousand deaths in the United States [1]. Governments world- wide have implemented similar policies to limit the spread of the virus, such as lock-down, social distance, mask policies. Research has proved the efficiency of the action behind each policy. For example, the use of masks was strongly protective, with a risk reduction of 70% for those that always wore a mask when going out. [2] Additionally, those infection rates are reduced drastically when social distancing intervention is implemented between 80% to 100%. [3] However, the actual effect of each policy was highly varied among countries due to factors such as socioeconomics [4,5]. Since issuing the policies with high efficiency to mitigate the pandemic are extremely important for all countries, we need to understand what factors affect the policy efficiency and how they affect it.

To solve these problems, previous studies have measured the effectiveness of several NPI (nonpharmaceutical interventions) [6,7,8], such as stay-at-home and business close policy, by measuring the change in real-time reproduction number (Rt), the expected number of new infections caused by an infectious individual in a population where some individuals may no longer be susceptible [9], and compare the decrease in real-time reproduction number after the policy is issued among different countries to see which country performs best for a specific policy.

In this study we used an eSIR model, extended state-space SIR models [10], to predict a one-month daily infection proportion after a period of lagging time and compare this to the actual daily infection proportion to measure the effectiveness policy.

Then, we conducted multivariate and single variate linear regression models, to study the correlation between policy effective rate and different potential factors, such as economics, population density, education level, etc. The results from this study can serve as a reference for governments when issuing covid polices

## 2. DATA

We obtained COVID data from the website covidtracking.com and the website collected and published the most complete data about COVID-19 in the US, including the daily death, cases, and hospitalization, etc.[11] This data source started collecting daily recovery rates, which are essential for the eSIR model prediction, in March 2020. Additionally, they are no longer collecting data as of March 7, 2021. Thus, the data used in this analysis are from March 2020 to March 2021.

We obtained data for the exact time when a policy was issued in each state from ballotpedia.org, which provides a detailed policy timeline for each state. [12] Data for policy effectiveness's potential factors are collected from various resources [13-21].

## 3. METHOD

We used the extended state-space SIR epidemiological model (eSIR)[10], to estimate the effectiveness of different policies and vaccine interventions on Covid-19. It uses the susceptible, infection, and removed as the input variables to predict the daily proportion of infection and proportion of death for a given amount of time. This approach has also been used during SARS.

The given dataset updated the cumulative recovered data every 7 days. To prevent the huge increase in recovered data every 7 days from impacting the model prediction, we use the loess.as() function under package (fANCOVA) to fit a smooth curve between all the recovered data. This function can choose a span value, the parameter α which controls the degree of smoothing, that optimizes the fit of the LOESS curve by fitting a LOESS regression and automates the parameter selection process.[22]   The smoothed recovered data was used for the rest of our analysis.

We first tested the prediction ability of eSIR model by setting a day when there was no major policy or vaccine issued 30 days before and after as the first prediction date and predict the daily infection proportion for the following 30 days. The eSIR model output an graph with infection proportion as the y axis, and date as x axis. It also includes two curves, one of which is the actual daily infection proportion and the other represents the predicted daily infection curved. The predicted curve is based on the medium values of predicted interval from eSIR model, since mediums are not affected by the outliers. Then, we compared the prediction curve and the actual daily infection proportion to see if the model can make accurate prediction on daily infection proportion when there was no policy and vaccine to interfere the result. We also compute the 90% or 95%? confidence interval.

For the policy intervention, we set the first prediction date 14 days after the policy was issued, because we assume that a 14-days lag time for counts of cases to coincide with the approximate incubation period of COVID-19.[6] While for the vaccine, according to CDC, the vaccine will be effective 14 days after the first shot, and we also set a 14-day lag time for vaccine. Thus, we set the first prediction date 28 days after the first shot. [23]

The effectiveness of policy intervention and vaccine is defined as the policy effective rate which is measured by the infection proportion prediction minus actual infection proportion and dived by of actual infection proportion. We calculated the daily policy effective rate and define the largest value among all 31 prediction days as max policy effective rate. We also calculate the sum of infection proportion prediction minus the sum of actual infection proportion and dived by the sum of actual infection proportion and defined it as the total policy effectiveness rate.

We use the total policy effective rate and max policy effective rate as measurements to compare the effectiveness of certain policies between states.

Additionally, we applied Inverse-normal transformation to both the total policy effective rate and max policy effective rate. Then, we test the multivariate linear model to figure out which factor has a significant impact on the effectiveness of each policy. We test factors such as public health funding per person, number of hospital beds per 1000 people, and GDP per capita, etc. Since we only have 21 or 28 states available for each policy, we set the level of significance as 0.1. We also conducted a single variate linear model between the max or total policy effective rate and each potential factor since the significance of each potential may decrease in a multivariate linear regression model because of collinearity between factors. The level of significance, in this case, was still set to be 0.1. We studied which factors significantly impact max or total policy effective rate based on the results from multivariate and single variate linear regression model results.

## 4. RESULT

### 4.1. MODEL EFFICIENCY:

The estimated daily infected proportion for state Alabama from 06/21/20 to 07/21/20 is based on the daily infection proportion from 05/22/20 to 06/21/20 as shown in figure 1. The horizontal axis displays the date month/day/year, and the vertical axis displays the proportion of infected. The green dotted line represents the actual daily infection proportion, while the red line represents the predicted values. The red range represents the confidence interval. The figure shows that the predicted daily infected rate is similar to the actual one and all the actual values fall in the 95% confidence interval. We applied the same method to all 26 states and get similar results. Thus we conclude that the eSIR is efficient in predicting accurate daily infection proportion when there is no policy interfered for all 26 states.

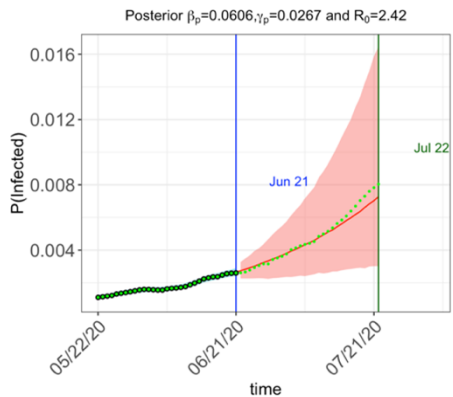

Figure 1.   Model Effectiveness Test figure. The x-axis represents infection proportion and the y-axis represents the date. The blue vertical line indicates the first prediction day, while the green line indicates the last prediction day. The green dot curve represents the actual daily infection proportion. The red line represents the predicted infected curve measured by the eSIR model with medians as the predicted value.

### *4.2. THE MASK PLOT:*

Figure 2 shows the predicted daily infected proportion for Alabama 14 days after the state-wide Mask policy was issued on 07/01/2020. The elements in the graph are the same as Figure 1. Figure 2 indicates that the predicted line is first lower than the actual value curve, but slowly catch up and pass over the actual value curve around the middle, and finally becomes much larger than the actual value curve, which indicates that the mask policy effect starts to show after a period of extra lagging time and the effect is significant.

We apply the same method to the states that have data available for the tested period and had issued a state-wide mask policy, which is 19 states in total. The mask policy plots have many variations among states. Most of the states, such as New Mexico, show an immediate decrease in the actual values curve compared to the predicted curve, which indicates that the mask policy has started to impact daily infected proportion after 14 days for the policy lagging period. Some of the states, such as Arkansas, show a similar pattern to Alabama mask policy, which shows impact after a short period of lagging. However, we also have some states, such as Vermont, which show no difference between the predicted curve and actual curve and thus show no impact for the mask policy. Also, some states, such as Michigan, indicate an opposite result, in which the predicted curve is smaller than the actual curve.

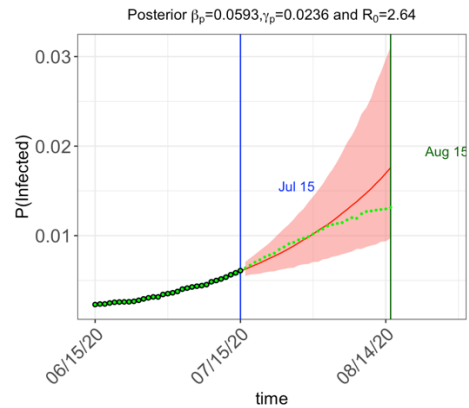

Figure 2.   Mask Policy effectiveness figure. The x-axis represents infection proportion and the y-axis represents the date. The blue vertical line indicates the first prediction day, which is 14 days after the mask policy is issued, while the green line indicates the last prediction day. The green dot curve represents the actual daily infection proportion. The red line represents the predicted infected curve measured by the eSIR model with medians as the predicted value.

### *4.3. THE VACCINE PLOT:*

Figure 3 shows the predicted daily infected proportion for Alabama 28 days after the Covid vaccine was first distributed 12/01/2020 for all states. The elements in the graph are the same as Figure 1 and Figure 2. Figure 2 indicates that the predicted curve is larger than the actual curve on the first day of the prediction and the difference kept increasing as time goes, which indicates that vaccine policy immediately shows an impact on daily infected proportion after 28 days of lagging.

We apply the same method to all the states that have data available for the test, which is in 26 states in total. The vaccine policy plots have much fewer variations among states compared to the mask policy. Most of the states, such as West Virginia, have a similar pattern as Alabama, which indicates that the mask policy has started to impact daily infected proportion. Some of the states, such as South Carolina, show impact after a short period of lagging. However, we also have some states, such as Maryland, which shows no difference between the predicted curve and actual curve and thus show not impact for the vaccine policy. Also, some states, such as Texas, indicate an opposite result, in which the predicted curve is smaller than the actual curve.

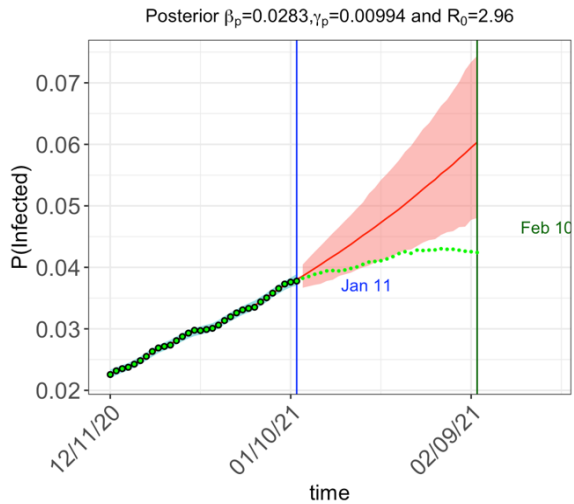

Figure 3.   Vaccine Policy Effectiveness Test figure. The x-axis represents infection proportion and the y-axis represents the date. The blue vertical line indicates the first prediction day, which is also 28 days after the vaccine is distributed, while the green line indicates the last prediction day. The green dot curve represents the actual daily infection proportion. The red line represents the predicted infected curve measured by the eSIR model with medians as the predicted value.

### 4.3 TOTAL POLICY EFFECTIVE RATE

The figure 4 shows the total policy effective rate and max policy effective rate of each state. The x-axis shows the state, and the y axis shows the total policy effective rate. The blue bars show the total policy effective rate for mask policy; the orange bars show the total policy effective rate for the vaccine; the grey bars show the max policy effective rate for mask policy, the yellow bars show the max policy effective rate for the vaccine. It is ordered by the total policy effective rate because all 26 states have distributed vaccines.   All the states in figure 4 distributed vaccines state-wide, while only 21 of them have issued mask policy and have data available for analysis. (Oklahoma has not issued a state-wide mask policy, and other states do not have recovery data available for analysis.)

In figure 4, we notice that the results for the total policy effective rate are consistent with the ones for the max policy effective rate. However, the total policy effective rates for mask policy tend to have more negative values, while the max policy effective rates for mask policy tend to have more extreme high values. It is because, in most of these states, the prediction curve increases much more quickly than the actual curve, which leads to a huge gap on the last day of prediction. Additionally, the states with a high mask policy effective rate also tend to have a high vaccine policy effective rate. This applies to both the total policy effective rate and the max policy effective rate. The correlation between mask total policy effective rate and vaccine total policy effective rate is 0.556. The correlation between mask max policy effective rate and vaccine max policy effective rate is 0.496. Both are high enough to be considered as a high correlation.

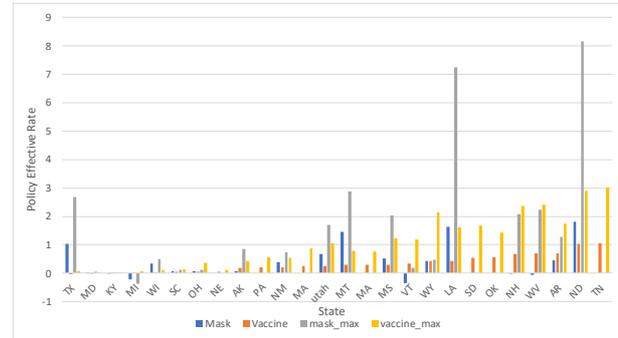

Figure 4.   Vaccine Policy Effectiveness Test figure. The x-axis represents the states and the y-axis represents the date. The blue bars show the total policy effective rate for mask policy; the orange bars show the total policy effective rate for the vaccine; the grey bars show the max policy effective rate for mask policy, the yellow bars show the max policy effective rate for the vaccine.

### 4.4 FIND FACTORS THAT AFFECT POLICY EFFECTIVE RATE BASED ON LINEAR REGRESSION MODEL

Based on the result above, we notice the variation within total and max policy effective rate between the states. We want to explore more about what factors lead to the difference in policy effectiveness among states. We use the multivariate linear model and single variate linear model to explore the impact of different potential factors.

#### 4.4.1   MULTIVARIATE LINEAR REGRESSION MODEL BETWEEN POLICY EFFECTIVE RATE DIFFERENT FACTORS

In table 1, we include two factors with the least low p-value in each policy effective rate's multivariate linear model. The corresponding estimated coefficient and p values are also listed in the table. All tables that include the estimated coefficient, standard errors, test statistics, and p values for each factor in each policy effective rate's multivariate linear model are in the appendix. It also shows the R squared value of the multivariate linear model of each policy effective rate.

In mask total policy effective rate and max policy effective rate multivariate linear models, only the population density is smaller than the level of significance, which indicates that public health per person has the most significant impact on the effectiveness of mask policy. Besides population density, the percent of the population over 65 also has a p-value relatively close to the level of significance.

In vaccine total policy effective rate and max policy effective rate multivariate linear models, only the public health funding per person has a p-value less than the level of significance 0.1, which indicates that public health per person has the most significant impact on the effectiveness of vaccine policy. Besides public health funding per person, immigration also has a relatively low p-value.

The R squared of the multivariate linear model for the max mask, total mask, and max vaccine policy effective rate are around 0.4. Although 0.4 is usually considered as a low correlation for the linear regression model, our dataset is so small that we still consider them as significant correlation. Additionally, the R squared of the multivariate linear model for the max mask for total vaccine policy effective rate is above 0.7, which indicates a strong correlation.

| policy.effective.rate | factor | estimate | p.value |
| --- | --- | --- | --- |
| max mask ($R^2$ = 0.3922) | population density | -0.00386 | 0.1373 |
|  | hospital bed | 0.4765 | 0.4977 |
| total mask ($R^2$ = 0.4786) | population density | -0.003272 | 0.07596 |
|  | precent of population over 65 | -0.2287 | 0.1074 |
| max vaccine ($R^2$ = 0.4365) | public health | 0.02445 | 0.07203 |
|  | immigration | -0.09829 | 0.334 |
| total vaccine ($R^2$ = 0.7319) | public health | 0.02914 | 0.02472 |
|  | immigration | -0.1154 | 0.2248 |

Table 1.   Multivariate linear regression model. The first column represents the type of policy effective rate and R^2 value of its multivariate linear regression model. The second column represents the two most significant factors for each policy effective factor according to its multivariate linear regression model. The third and fourth columns represent each factor's estimated coefficient and p-value.

### 4.4.2 SINGLE VARIATE LINEAR REGRESSION MODEL BETWEEN POLICY EFFECTIVE RATE DIFFERENT FACTORS

Table 2 shows all the factors that have a p-value less than 0.1 level of significance for each policy effective rate. It also includes the estimated coefficient and p-value of each factor.

For mask policy effective rate, we have population density, voters in each state predominantly choose either the Republican Party or Democratic Party, the number of hospital beds per 1000 people, and percent of the population over 65 have the most important impact on the effectiveness of mask policy.

For vaccine policy effective rate, we have percent of immigration in each state, the number of hospital beds per 1000 people, and public health funding per person have the most important impact on the effectiveness of mask policy.

| policy.effective.rate | factor | estimate | p.value |
| --- | --- | --- | --- |
| max mask | population density | -0.003014 | 0.0572 |
|  | politics (R) | 0.8419 | 0.0857 |
|  | hospital bed | 0.5753 | 0.0591 |
| total mask | population density | -0.003032 | 0.0555 |
|  | politics (R) | 1.093 | 0.0208 |
|  | percent of population over 65 | -0.2527 | 0.0182 |
|  | hospita bed | 0.5235 | 0.0889 |
| max vaccine | immigration | -0.08807 | 0.0529 |
|  | hospital bed | 0.5426 | 0.0332 |
| total vaccine | immigration | -0.1027 | 0.0218 |
|  | hospital bed | 0.556 | 0.0286 |
|  | public health | 0.02024 | 0.0547 |

Table 2.   Single variate linear regression model. The first column represents the type of policy effective. The second column represents factors with p-value less than 0.1 for each policy effective factor according to its multivariate linear regression model. The third and fourth columns represent each factor's estimated coefficient and p-value.

## 5  DISCUSSION

Based on the two tables above, we conclude that population density and percent of the population over 65 have a negative impact on the mask policy effective rate. According to CDC, transmission can be reduced by up to 96.5% if both an infected person and an uninfected person wear tightly fitted surgical masks or a cloth mask together with a surgical mask. [25] However, unlike surgical masks, cloth masks' ability to block transmission is highly variable due to their design, fit, and materials used [26]. Thus, it is possible that the spreading of covid-19 will be efficiently reduced with low population density even only cloth masks are used, but with high population density, cloth masks may not decrease the spreading of covid 19 very efficiently, due to their lower blocking rate. For the negative impact from percent of the population over 65, it is possibly because old people on average require more time to recover from Covid-19 [27] and thus have a much longer period of hospitalization. A higher old population percent means more proportion of the infected population will also be old people. Since the infection proportion is calculated by the total infected population minus the removed population (removed and recovered population), longer hospitalization for the elder population will result in a smaller removed population and thus a larger infection proportion and lower policy effective rate.

However, voters in each state predominantly choose Republican Party, and the number of hospital beds per 1000 people have a positive impact on the mask policy effective rate. Most of the states first issued their mask policy during July when Republican-led states had a higher positive-testing and COVID-19 case diagnoses than democracy-led states overall. [28] Thus, higher infection proportion before mask policy is issued might at least partially explain why Republican-led states have a larger gap between prediction curve and actual curve have which leads to a higher effective rate. With the decrease in infection, states with a larger number of hospital beds per 1000 people will be recovered from 111e. Thus, these states have a higher mask policy effective rate.

For vaccine policy effective rate, the number of hospital beds per 1000 people and public health funding per person have a positive impact on it. The positive impact from the number of hospital beds per 1000 people will be likely due to similar reasons above. Additionally, vaccination will mitigate the symptom and recovery time. States with higher public health funding per person will be more likely to distribute the vaccine better, as they may have more vaccines available overall, so they have a higher vaccine policy effective rate. However, percent of immigration has a negative impact on the vaccine policy effective rate. A possible reason for that is vaccine sites across the U.S. require some form of identification, a requirement that many undocumented immigrates do not have, so the vaccine is less efficient for states with a higher immigration population. [29]

# 6 CONCLUSION

The current research was designed to examine the effectiveness of different policies. Significant differences in policy effective rates of the same policy for different states have been identified. Several reasons may help interpret these findings such as the positive impact from the number of hospital beds per 1000 people, public health funding per person, and a negative impact from percent of immigration for vaccine policy effective rate; additionally, the positive impact from the number of voters in each state predominantly choosing Republican Party, and the number of hospital beds per 1000 people, and a negative impact from population density and percent of the population over 65 for mask policy effective rate.

This research uses a new method to sheds light on the difference in policy effective rate between states of the same policy and the correlation between policy effective rate and different factors. It can be served as a reference for future covid policy.

Due to the lack of enough data, this research only includes 26 states for vaccine policy and 19 states for mask policy, which negatively affects the significance of all the potential factors. If the data for all 55 states are available, the linear model will be improved, and it is more likely to find more significant factors that impact the policy effective rate. Also, the prediction curve in this research is based on the eSIR model. Although it has been proved to be an effective model for predicting short-range infection proportion, the prediction is not a hundred percent accurate. It is still possible that the difference between the prediction curve and the actual curve is partly due to the uncorrected prediction by the eSIR model. The results will be more promising if a linear regression model is applied to the change in real-time reproduction number and get similar results. Additionally, this research only computed the policy effective rate based on infection proportion. However, the vaccine has also been proved to reduce the death rate, so redoing all the processes based on death proportion is a possible choice for future research.

**REFERENCES**

[1]  Johns Hopkins Coronavirus Resource Center. (n.d.). Covid-19 United States cases by county. COVID-19 United States Cases by County. Retrieved October 28, 2021, from https://coronavirus.jhu.edu/us-map..

[2]  Wu J, Xu F, Zhou W, Feikin DR, Lin CY, He X, Zhu Z, Liang W, Chin DP, Schuchat A. Risk factors for SARS among persons without known contact with SARS patients, Beijing, China. Emerg Infect Dis. 2004 Feb;10(2):210-6. doi: 10.3201/eid1002.030730. PMID: 15030685; PMCID: PMC3322931.

[3]  Daghriri T, Ozmen O. Quantifying the Effects of Social Distancing on the Spread of COVID-19. Int J Environ Res Public Health. 2021 May 23;18(11):5566. doi: 10.3390/ijerph18115566. PMID: 34071047; PMCID: PMC8197116.

[4]  Morgan Pincombe, Victoria Reese, Carrie B Dolan, The effectiveness of national-level containment and closure policies across income levels during the COVID-19 pandemic: an analysis of 113 countries, Health Policy and Planning, Volume 36, Issue 7, August 2021, Pages 1152–1162, https://doi.org/10.1093/heapol/czab054.

[5]  Rabail Chaudhry, George Dranitsaris, Talha Mubashir, Justyna Bartoszko, Sheila Riazi, A country level analysis measuring the impact of government actions, country preparedness and

socioeconomic factors on COVID-19 mortality and related health outcomes, EClinicalMedicine, Volume 25, 2020, 100464, ISSN 2589-5370,

https://doi.org/10.1016/j.eclinm.2020.100464Tools , Entrepreneurship Research Journal, Published by De Gruyter March 13, 2015.

[6] Liu S, Ermolieva T, Cao G, Chen G, Zheng X. Analyzing the Effectiveness of COVID-19 Lockdown Policies Using the Time-Dependent Reproduction Number and the Regression Discontinuity Framework: Comparison between Countries. Engineering Proceedings. 2021; 5(1):8. https://doi.org/10.3390/engproc2021005008

[7] Inglesby TV. Public Health Measures and the Reproduction Number of SARS-CoV-2. JAMA. 2020;323(21):2186–2187. doi:10.1001/jama.2020.7878.

[8] Arroyo-Marioli F, Bullano F, Kucinskas S, Rondón-Moreno C (2021) Tracking R of COVID-19: A new real-time estimation using the Kalman filter. PLoS ONE 16(1): e0244474. https://doi.org/ 10.1371/journal.pone.0244474.

[9] Gostic KM, McGough L, Baskerville EB, et al. Practical considerations for measuring the effective reproductive number, Rt. Preprint. medRxiv. 2020;2020.06.18.20134858. Published 2020 Aug 28. doi:10.1101/2020.06.18.20134858.

[10] lilywang1988. (2019). Extended state-space sir epidemiological models. GitHub. Retrieved November 8, 2021, from https://github.com/lilywang1988/eSIR.

[11] The Atlantic Monthly Group. (2021). The Data. The COVID Tracking Project. Retrieved November 8, 2021, from https://covidtracking.com/data.

[12] State-level mask requirements in response to the coronavirus (COVID-19) pandemic, 2020-2021. Ballotpedia. (n.d.). Retrieved November 9, 2021, from https://ballotpedia.org/State-level_mask_requirements_in_response_to_the_coronavirus_(COVID-19)_pandemic,_2020-2021.

[13] For GDP by state, see "Gross domestic product (GDP) by state (millions of current dollars)". Bureau of Economic Analysis. Retrieved 8 June 2017.

[14] For population by state, see "Annual Estimates of the Resident Population for the United States, Regions, States, and Puerto Rico: April 1, 2010 to July 1, 2016" (XLSX). United States Census Bureau. Retrieved 8 June 2017.

[15] US NEWS. (n.d.). Most educated states 2021. Retrieved November 8, 2021, from https://worldpopulationreview.com/state-rankings/most-educated-states.

[16] Wikimedia Foundation. (2021, October 14). Political party strength in U.S. states. Wikipedia. Retrieved November 8, 2021, from https://en.wikipedia.org/wiki/Political_party_strength_in_U.S._states#cite_note-NJvoterreg-32.

[17] U.S. Census Bureau (2019). Nativity in the United States American Community Survey 1-year estimates. Retrieved from https://censusreporter.org

[18] Published by Statista Research Department, & 28, O. (2021, October 28). Senior population of the U.S. by State 2019. Statista. Retrieved November 8, 2021, from https://www.statista.com/statistics/736211/senior-population-of-the-us-by-state/.

[19] KFF. (2021, April 5). Hospital beds per 1,000 population by ownership type. KFF. Retrieved November 8, 2021, from https://www.kff.org/other/state-indicator/beds-by-ownership/?currentTimeframe=0&sortModel=%7B%22colId%22%3A%22Location%22%2C%22sort%22%3A%22asc%22%7D#notes.

[20] SHADAC analysis of [Per person state public health funding], State Health Compare, SHADAC, University of Minnesota, statehealthcompare.shadac.org, Accessed [11/08/2021]. http://statehealthcompare.shadac.org/rank/117/per-person-state-public-health-funding#2,3,4,5,6,7,8,9,10,11,12,13,14,15,16,17,18,19,20,21,22,23,24,25,26,27,28,29,30,31,32,33,34,35,36,37,38,39,40,41,42,43,44,45,46,47,48,49,50,51,52/a/32/154/false/location

[21] Published by Statista Research Department, & 21, J. (2021, January 21). Population density in the U.S., by State 2020. Statista. Retrieved November 8, 2021, from https://www.statista.com/statistics/183588/population-density-in-the-federal-states-of-the-us/.

[22] Wang, X., & Ji, X. (2020, November 13). Nonparametric analysis of covariance [R package fANCOVA version 0.6-1]. The Comprehensive R Archive Network. Retrieved November 10, 2021, from https://cran.r-project.org/web/packages/fANCOVA/index.html.

[23] Kriss JL, Reynolds LE, Wang A, et al. COVID-19 Vaccine Second-Dose Completion and Interval Between First and Second Doses Among Vaccinated Persons — United States, December 14, 2020−February 14, 2021. MMWR Morb Mortal Wkly Rep 2021;70:389–395. DOI:

http://dx.doi.org/10.15585/mmwr.mm7011e2external icon.

[24] Cori, A.; Ferguson, N.M.; Fraser, C.; Cauchemez, S. A New Framework and Software to Estimate Time-Varying Reproduction Numbers During Epidemics. Am. J. Epidemiol. 2013, 178, 1505–1512.

[25] Rabin, R. C., & Stolberg, S. G. (2021, February 10). The C.D.C. says tight-fit masks or double masking with cloth and surgical masks increases protection. The New York Times. Retrieved November 19, 2021, from https://www.nytimes.com/2021/02/10/world/double-mask-protection-cdc.html.

[26] Brooks JT, Butler JC. Effectiveness of Mask Wearing to Control Community Spread of SARS-CoV-2. JAMA. 2021;325(10):998–999. doi:10.1001/jama.2021.1505

[27] Voinsky I, Baristaite G, Gurwitz D. Effects of age and sex on recovery from COVID-19: Analysis of 5769 Israeli patients. J Infect. 2020;81(2):e102-e103. doi:10.1016/j.jinf.2020.05.026

[28] Kempler, C. (2021, March 10). As cases spread across U.S.. Last year, pattern emerged suggesting link between governors' party affiliation and covid-19 case and death numbers. Johns Hopkins Bloomberg School of Public Health. Retrieved November 19, 2021, from https://publichealth.jhu.edu/2021/as-cases-spread-across-us-last-year-pattern-emerged-suggesting-link-between-governors-party-affiliation-and-covid-19-case-and-death-numbers.

[29] 29. Semotiuk, A. J. (2021, September 2). Immigrants slow to get covid-19 vaccine as impediments block their access. Forbes. Retrieved November 19, 2021, from https://www.forbes.com/sites/andyjsemotiuk/2021/09/01/immigrants-slow-to-get-covid-19-vaccine-as-impediments-block-their-access/?sh=1aedc5ed5333.